\documentstyle[11pt]{article}
\catcode`\@=11
\begin{document} \openup6pt

\title{  Gravitational Field of A Radiating Star in Higher Dimensions }

\author{B. C. Paul\thanks{Electronic mail : bcpaul@iucaa.ernet.in}  \\
	Physics Department, North Bengal University, \\
Siliguri, Dist. : Darjeeling, Pin : 734 430, West Bengal, India \\
P. K. Chattopadhyay \\
Physics Department, Alipurduar College \\ PO : Alipurduar Court, Jalpaiguri, Pin : 736122, India}

\date{}

\maketitle

\vspace{0.5in}

\begin{abstract}

We obtain fields of a relativistic radiating star of non-static mass in the framework of higher dimensional spacetime. Assuming    energy-momentum tensor in Higher dimensions analogous to that considered by  Vaidya in 4 dimensions we obtain solution of a radiating spherically symmetric star.
The solution obtained here is  new in higher dimensions which however reduces to that obtained by Vaidya in 4 dimensions. It is also different in form from that obtained by Iyer and Vishveshwara.
The interesting observation is that the  radius of a radiating star in higher dimensions   oscillates. The radial size of radiating star oscillates with a  period which depends on the modes of vibration
and dimensions of the space-time.

\end{abstract}


\vspace{0.2cm}

\vspace{4.5cm}

\pagebreak

\section{ Introduction }

Kaluza and Klein [1] first tried to unify gravity with electromagnetism by introducing an extra dimensions. The early attempt could not work well.  Recently the approach has been revived  and considerably generalized after realizing that many interesting theories of particle interactions need space-time dimensions more than the usual four for their consistent formulation. It is now believed that superstring theory which is consistent in 10 dimensions may be a promising candidate where all the forces in nature may be unified. As  the extra dimensions are not observable there were various proposals put forwarded to resolve the issue. The present idea is that our cosmos may be a 3-brane evolving in a D-dimensional space-time [2]. It is also proposed that a large number of extra dimensions [3] of the spacetime in the scenario may be accommodated.      Cadeau and Woolgar [4] addressed the issue in the context of black holes which led to homogeneous but non-FRW -braneworld cosmologies. There has been a  growing interest in recent years in obtaining a higher dimensional analogue of a four dimensional general relativistic results because of the successes of Superstring/M- theory. Several works in the literature have appeared which include the higher -dimensional generalization of the spherically symmetric Schwarzschild and Reisner-Nordstr$\ddot{o}$m black holes [5, 6], Kerr black holes [7], Vaidya solution [8], generalization of mass to radius ratio of a uniform density star [9]. Mandelbrot [10] studied the problem on the variability of dimensions in which he describes how a ball of thin thread is seen as an observer changes scale. An object which seems to be a point object from far point becomes a three-dimensional ball at  closer distances. Thus, as an observer moves down through various scales the ball appears to keep changing its shape. In this case although the effective dimension of the contents in the space changes but the embedding dimensions of the ball do not change. It is possible   that there are compact [2] or non-compact [3, 11] dimensions present at a certain point. At this scale, the (3 +1) metric is simply not true, although one obtains a valid description with general relativity. Liu {\it et al.} [12] reported solar system tests based on a five dimensional extension of the Schwarzschild metric and Cassisi  {\it et al.}  [13] have examined the effects of higher dimensions on stellar evolution. Yu and Ford [11] reported that observable effects of higher dimensions may be found from lightcone fluctuations. Guenther and Zhuk [14] investigated the observable consequences of spontaneous compactification hypothesis for the extra dimension. At present, dimensional physics has become an active area of investigation with some promise of future experimental insights [15].

Recently, works of Vaidya [16-18] in 4 dimensions which generalizes  the  static Schwarzschild's solution incorporating  a  non-static mass  in General Relativity have been re-printed  in General Relativity and  Gravitation Journal to focus the importance of his works  done in 1951 .  
Prior to 1951, there were various atempts [19]  to generalize the  static Schwarzschild's solution to incorporate a  non-static mass and radiating  solution in General Relativity. However,  Vaidya [18] first obtained a generalized solution of a radiating star.
The components of the corresponding energy-momentum tensor  of a radiating star and the properties of some of its quantities related to a radiating mass of star had been studied in his famous papers [16-18].
 In this paper we intend to explore the work of Vaidya [18]  in   higher dimensional framework considering a D-dimensional space-time. We determine the  instantaneous  radius of the
radiation zone of the star assuming the fact that the  mass inside the radiation zone is  a function of both  $r$ and $t$ ($M = M(r,t)$). It is also assumed that beyond the bounding sphere of the radiation zone, the space is considered to be empty and the corresponding metric  is  described by Schwarzschild static solution in higher dimensions [20]. The mass parameter arising in this case  is related to  the space-time
dimensions. 

The paper is organized as follows: In section 2, we set up the relevant field equations of a radiating star in higher dimensions and introduce  the corresponding energy momentum tensor.  
In section 3, we present solution of radiating relativistic star, its physical properties, instantaneous boundary of the radiation zone. Finally, we presemt  a brief discussion in section 4.

\vspace{0.5cm}

\section{  Field Equation of a Radiating Star in Higher Dimension}

The Einstein's field equation in higher dimension is given by
\begin{equation}
R_{\mu\nu} - \frac{1}{2}g_{\mu\nu} R = 8\pi G_D T_{\mu\nu}
\end{equation}
where D is the total number of dimensions and Greek indices $\mu, \; 
\nu = (0, 1, 2, .....D)$, $G_D = G
 V_{D-4} $ is the gravitational constant in $D$ dimensions,
$G$ denotes the 4 dimensional Newton's constant and $V_{D-4}$ is the volume of the extra dimension, $R_{\mu\nu}$ is the
Ricci tensor and $T_{\mu\nu}$ is the energy momentum tensor.  The line element of a higher dimensional spherically symmetric non-static space time is
\begin{equation}
 ds^2 = e^{\nu(r, t)} dt^2 - e^{\mu(r, t)}  dr^2 - r^2 d\Omega_n^2
\end{equation}
where $\mu$ and $\nu$ are functions of $r$ and $t$, $ n = D-2$ and $d\Omega_n^2 = d\theta_1^2 + sin^2 \; \theta_1 d\theta_2^2 +
sin^2\theta_1 \; sin^2\theta_2 \;  d\theta_3^2+ ........
+(sin^2\theta_1 \; sin^2\theta_2  ..... sin^2\theta_{n-1} \; d\theta_n^2)$
represents the metric on the $n$-sphere in polar co-ordinates. Using the metric (2) in  the  field equation (1) we get the following  relevant equations :
study of the following five equations:
\begin{equation}
8 \pi G_D T_0^{0} =  \frac{n \mu' e^{-\mu}}{2r} + \frac{n(n-1)}{2r^2} (1- e^{-\mu}) ,
\end{equation}
\begin{equation}
8 \pi G_D T_1^{1} = - \frac{n \nu' e^{-\mu}}{2r} + \frac{n(n-1)}{2r^2} (1- e^{-\mu}) , 
\end{equation}
\[
8 \pi G_D T_2^{2} = 8 \pi G_D T_3^{3}= 8 \pi G_D T_4^{4} = .... = - \frac{1}{2} e^{-\mu}\left[\nu{''} + \frac{\nu{'}^{2}}{2} -
\frac{\mu{'} \nu{'}}{2} +\frac{(n-1)(\nu{'}-\mu{'})}{r}\right]  
\]
\begin{equation}
+ \frac{1}{2}e^{-\nu}\left[\ddot{\mu} + \frac{\dot{\mu}^{2}}{2}
- \frac{\dot{\mu} \dot{\nu}}{2}\right]+ \frac{(n-1)(n-2)}{2r^2} (1- e^{-\mu}) , 
\end{equation}
\begin{equation}
8 \pi G_D T_1^{0} = \frac{n\dot{\mu}{e^{-\nu}}}{2r}, 
\end{equation}
\begin{equation}
8 \pi G_D T_0^{1} = - \frac{n\dot{\mu}{e^{-\mu}}}{2r}. 
\end{equation}
Here we adopt the convention $c = 1$ and overhead dash or dot denoting derivative w.r.t. $r$ or $t$ respectively. Knowing the
component of $T^{\mu\nu}$ in one co-ordinate system we can find out them in the other co-ordinate system using the tensor
transformation
\begin{equation}
 T^{\mu\nu} = \frac{\partial x^\mu}{\partial x_0^\alpha} \frac{\partial x^\nu}{\partial x_0^\beta} T_0^{\alpha\beta}. 
\end{equation}
As the radiant energy travels along null-geodesics, for a general co-ordinate system with metric given by 
\begin{equation}
 ds^2 = g_{\mu\nu} \; dx^\mu \; dx^\nu
\end{equation}
one obtains that along the flow of radiation
\begin{equation}
 g_{\mu\nu}\;  dx^\mu \;  dx^\nu = 0. 
\end{equation}
In analogy to  the energy momentum components adopted in {\it Ref.} [17] in four dimensions we consider here the energy-momentum components in higher dimensions as
 $T_0^{00} = T_0^{11} = T_0^{01} = T_0^{10} = \rho$. Using above components  we rewrite eq.(8) which is given by
\begin{equation}
 T^{\mu\nu} = \left[\frac{\partial x^\mu}{\partial x_0^1} \frac{\partial x^\nu}{\partial x_0^1}+
\frac{\partial x^\mu}{\partial x_0^0} \frac{\partial x^\nu}{\partial x_0^0}+
\frac{\partial x^\mu}{\partial x_0^1} \frac{\partial x^\nu}{\partial x_0^0}+
\frac{\partial x^\mu}{\partial x_0^0} \frac{\partial x^\nu}{\partial x_0^1}\right]\rho. 
\end{equation}
One can  write
\begin{equation}
\frac{dx^\mu}{d\tau} = \frac{\partial x^\mu}{\partial x_0^a} \frac{dx_0^a}{d\tau}
\end{equation}
using $ dx_0^1 = dx_0^0 = d\tau (say) $. Thus, eq. (12) becomes
\begin{equation}
\frac{dx^\mu}{d\tau} = \frac{\partial x^\mu}{\partial x_0^1} + \frac{\partial x^\mu}{\partial x_0^0}. 
\end{equation}
Using eq. (13) in  eqs. (10) and (11), we get the following equations
\begin{equation}
 g_{\mu\nu} \; \frac{dx^\mu}{d\tau} \frac{dx^\nu}{d\tau} = 0 
\end{equation}
and
\begin{equation}
T^{\mu\nu} =  \rho \; \frac{dx^\mu}{d\tau} \frac{dx^\nu}{d\tau}.
\end{equation}
As $ T^{\mu\nu} = \rho\eta^\mu\eta^\nu $,  we can change the upper index  to lower  index multiplying  by  $g_{\mu\nu}$,
 consequently we get
\begin{equation}
T_\mu^\nu = \rho \; \eta_\mu\eta^\nu
\end{equation}
where $\eta^\mu = \frac{dx^\mu}{d\tau} $ and  $\eta^\nu = \frac{dx^\nu}{d\tau}$.
For the lines of flow along the  null-geodesics one has
\begin{equation}
\eta_\mu\eta^\nu = 0.
\end{equation}
The different components of the energy momentum tensor are given by 
\[ T_0^0 = \rho \; \;  \eta_0\eta^0, \;  \;  T_1^1 = \rho \; \eta_1\eta^1, 
\; \;  T_1^0 = \rho \; \eta_1\eta^0, 
\]
\begin{equation}
 T_0^1 = \rho \; \eta_0\eta^1,
\; \; 
T_2^2 = 0,  \; \; T_3^3 = 0. 
\end{equation}
However, for radial flow one has  $\eta^2 = 0$, $\eta^3 = 0$ and we get from 
eq.(17) 
\begin{equation}
 \eta_0\eta^0 + \eta_1\eta^1 + \eta_2\eta^2 + \eta_3\eta^3 + ....... = 0 . 
\end{equation}
Since $\eta_2\eta^2 = \eta_3\eta^3 = ...... = 0$ we get,
 $$\eta_0\eta^0 + \eta_1\eta^1 = 0. $$
We convert lower index to upper index using $ g_{\mu \nu}$, and  obtain 
\begin{equation}
   e^{\nu}{(\eta^0)}^2 - e^\mu{(\eta^1)}^2 = 0. 
\end{equation}
Now, eqs.(18)-(20) along with $g_{00}$ and $g_{11}$  gives the following equation:
\begin{equation}
T_1^0e^{(\nu-\mu)/2} + T_0^0 = 0.
\end{equation}
Using eqs, (3) and (7), the above equation yields 
\begin{equation}
\frac{n \mu' e^{-\mu}}{2r} + \frac{n(n-1)}{2 r^2} \left(1- e^{-\mu} \right) +  \frac{n \dot{\mu}}{2 r} e^{- \frac{(\mu+\nu)}{2}} 
 = 0. 
\end{equation}

Again we have
\begin{equation}
 T_0^0 + T_1^1 = 0, 
\end{equation}
cosequently from eqs. (3) and (4),  we get
\begin{equation}
\frac{ne^{-\mu}}{2}\left[\frac{(\mu'-\nu')}{r} - \frac{2(n-1)}{r^2}\right]+\frac{n(n-1)}{r^2} = 0. 
\end{equation}
Using the energy momentum components given by  eq. (18) in eq. (5) we further obtain
\[
-e^{-\mu}\left[\nu{''} + \frac{\nu{'}^{2}}{2} -
\frac{\mu{'} \nu{'}}{2} +\frac{(n-1)(\nu{'}-\mu{'})}{r}\right]+ e^{-\nu}\left[\ddot{\mu} + \frac{\dot{\mu}^{2}}{2}
- \frac{\dot{\mu} \dot{\nu}}{2}\right]
\]
\begin{equation}
 + \frac{(n-1)(n-2)}{r^2} (1- e^{-\mu})=0. 
\end{equation}
Inside the radiation zone of a radiating star,  the  higher dimensional analog of the Schwarzschild metric [20] 
is given by
\begin{equation}
 ds^2 = \left(1-\frac{c}{r^{n-1}}\right)dt^2 - \left(1-\frac{c}{r^{n-1}}\right)^{-1}dr^2 - r^2 \; d\Omega_n^2
\end{equation}
where $c = c(r,t)$ is related to the mass of a higher dimensional star which is also a function of 
$r $ and  $t$. 
However, the space is empty out side the region of
the radiation zone  and therefore we consider the following  metric   
\begin{equation}
ds^2 = \left(1-\frac{C}{r^{n-1}}\right)dt^2 - \left(1-\frac{C}{r^{n-1}}\right)^{-1}dr^2 - r^2 \; d\Omega_n^2
\end{equation}
where the capital $C$ is the value of $c$ (used in (26)) in the empty space which is a constant. However,  $C$  is now related to the total mass $M$ of the
star. $M$ is considered to be the initial mass of the star just when the star starts radiation or shining. The functional relation
between $C$ and $M$ in higher dimensions is given by [20]
$$ M = \left[\frac{nA_nC}{16\pi G_D}\right]$$
 where $n = D-2$ and
$A_n = \frac{2\pi^{(n+1)/2}}{\Gamma((n+1)/2)}$.\\
Comparing metrics given by (2) and (26) we get
\begin{equation}
 e^{-\mu} = \left(1-\frac{c}{r^{n-1}}\right); \; \;  c=c(r,t)
\end{equation}
from which $c$ can be obtained as
$$  c = r^{n-1}(1-e^{-\mu}).$$
Differentiating the above expression w.r.t. $r$ and $t$ and using eq. (22) we get
$$e^{-\mu/2}\frac{\partial c}{\partial r}+e^{-\nu/2}\frac{\partial c}{\partial t}=\frac{2r^ne^{-\mu/2}}{n}\left[
\frac{n \mu' e^{-\mu}}{2r} + \frac{n(n-1)}{2r^2} (1- e^{-\mu}) + \frac{n\dot{\mu}}{2r}e^{-(\mu+\nu)/2}\right]$$
which further reduces to 
\begin{equation}
e^{-\mu/2}\frac{\partial c}{\partial r}+e^{-\nu/2}\frac{\partial c}{\partial t} = 0. 
\end{equation}
Finally after simplification we get
\begin{equation}
e^{\nu/2}=-\frac{\dot c}{c'}\left(1-\frac{c}{r^{n-1}}\right)^{-\frac{1}{2}}. 
\end{equation}
Finally, using the above equation in  eq. (22) we get
\begin{equation}
 \frac{c'}{\dot c} \left(\frac{\dot c'}{c'}-\frac{\dot{c}}{c'^2} c''\right)= \frac{c(n-1)/r^n}{\left(1-\frac{c}{r^{n-1}}\right)}, 
\end{equation}
which can be rewritten as 
\begin{equation}
\frac{1}{x} \frac{dx}{dr} = \frac{dz}{z} 
\end{equation}
where $x = \frac{\dot{c}}{c'}$ and $z= \left( 1-\frac{c}{r^{n-1}} \right)$.
Integrating the above equation we get,
\begin{equation}
\frac{\dot c}{c'}=\left(1-\frac{c}{r^{n-1}}\right).
\end{equation}
It may be mentioned here that the above differential  equation is of different form from that obtained by Iyer and Visveshwara [8].
Thus we obtain the corresponding metric for an envelope of a radiating star in higher dimension ($D$) which is given by
\[
ds^2=\frac{\dot c^2}{c'^2}\left(1-\frac{c}{r^{n-1}}\right) \; dt^2 - \left(1-\frac{c}{r^{n-1}}\right)^{-1}dr^2-
r^2(d\theta_1^2 + sin^2\theta_1 d\theta_2^2 + sin^2\theta_1 sin^2\theta_2 d\theta_3^2 + 
\]
\begin{equation}
 ........+ (sin^2\theta_1sin^2\theta_2 .....sin^2\theta_{n-1} d\theta_n^2)) 
\end{equation}
with $\dot c = c' \left(1-\frac{c}{r^{n-1}}\right).$\\
The surviving components of the energy-momentum tensor are
\begin{equation}
T_0^0 = \frac{nc'}{16\pi G_D r^n} \; \;  T_1^1 = -\frac{nc'}{16\pi G_D r^n} \; \;  T_0^1 = -\frac{\dot c}{8\pi G_D r^n}\; \;  T_1^0 = \frac{c'^2}{8\pi G_D\dot c r^n}
\end{equation}
from the above expression we see that the components of the energy-momentum tensor now depends  
on dimensions of the universe too. In four dimension  i.e. $ n=2 $ and puting  $c=2m$  and $G=1$. We recover the
corresponding energy-momentum tensor that has been obtained by Vaidya [18]. The energy momentum tensors are as follows 
\begin{equation}
T_0^0 = \frac{m'}{4\pi r^2} \; \;  T_1^1 = -\frac{m'}{4\pi r^2} \; \;  T_0^1 = -\frac{\dot m}{4\pi r^2}
\; \;  T_1^0 = \frac{m'^2}{4\pi \dot m r^2}
\end{equation}
Let us now calculate the values of certain quantities related to the radiating star in higher
dimensions. Previously we have defined a quantity $ \eta^\mu $,  where $\mu = 0, 1, 2, ... , (D-1)$.

But as the lines of flow are null-geodesics in this case  one considers $\eta^2=\eta^3=....=0$. Therefore our
task is to determine $\eta^0 $ and $\eta^1$. Let us now define an operator with $\eta^0 $ and $\eta^1$
which is
$$\frac{d}{d\tau}=\eta^0\frac{\partial}{\partial t}+\eta^1\frac{\partial}{\partial r}.$$
Applying the operator on  eq.(28) we get,
\begin{equation}
\frac{dc}{d\tau}= 0.
\end{equation}
As the lines of flow are null-geodesic, eliminating $\eta^0$ from the relations $\eta_\mu\eta^\mu=0$ and
$(\eta^\mu)_\nu\eta^\nu=0$ we get,
\begin{equation}
\frac{\partial \eta^1}{\partial r}+\frac{\partial \eta^1}{\partial t} \; e^{(\mu-\nu)/2}+\eta^1\left[\frac{\mu'+\nu'}{2}
+\dot \mu e^{(\mu-\nu)/2}\right]=0. 
\end{equation}
and eliminating  $\eta^0$  from $\eta_\mu\eta^\mu=0$ and $(T^{\mu\nu})_\nu=0$ we get,
\begin{equation}
\frac{\partial}{\partial r} \left(\rho r^2\eta^1 \right)+e^{(\mu-\nu)/2} \; \frac{\partial}{\partial t} \left(\rho r^2\eta^1 \right)
+ \left(\rho r^2\eta^1 \right) \left[\frac{\mu'+\nu'}{2}+\dot \mu \; e^{(\mu-\nu)/2}\right]=0
\end{equation}
The connection between $\mu'$ and $\nu'$ can be obtained from eqs. (22) and (24) which is given by
$$ \left[\frac{\mu'+\nu'}{2}+\dot \mu e^{(\mu-\nu)/2}\right]=0.$$
Thus  eqs. (38) and (39) may now be written as 
\begin{equation}
 \frac{\partial \eta^1}{\partial r}+\frac{\partial \eta^1}{\partial t}e^{(\mu-\nu)/2}=0, 
\end{equation}
\begin{equation}
\frac{\partial}{\partial r}(\rho r^2\eta^1)+e^{(\mu-\nu)/2}\frac{\partial}{\partial t}(\rho r^2\eta^1)=0. 
\end{equation}
The above two equations leads to the following relations 
\begin{equation}
 \frac{d\eta^1}{d\tau}= 0,
\end{equation}
\begin{equation}
 \frac{d}{d\tau}(\rho r^2\eta^1)= 0.
\end{equation}
Finally we get
\begin{equation}
\frac{d}{d\tau}(\rho r^2)= 0
\end{equation}
It is evident from eqs. (37) ,(42) and (44) that the  quantities  c, $\eta^1$,$\rho r^2$ are conserved along the 
flow of radiation in the case
of higher dimensional radiating star also. Let us now determine the actual values of $\eta^0$ and $\eta^1$
in the case of a radiating star in higher dimensions. From eq. (40) we get,
$$\frac{\partial \eta^1}{\partial r}+\frac{\partial \eta^1}{\partial t}e^{(\mu-\nu)/2}=0,$$
With the help of eq.(29),  the above equation can be written as 
\begin{equation}
\frac{\partial \eta^1}{\partial r} - \frac{c'}{\dot c}\frac{\partial \eta^1}{\partial t}=0.
\end{equation}
We denote here
\begin{equation}
\eta^1=\psi(c)
\end{equation}
which enable us to write eq.(20)  as
\begin{equation}
\eta^0 = -\frac{c'}{\dot c}\psi(c)
\end{equation}
and the non-zero  off diagonal  component of the   energy momentum tensor is
$$T_0^1=-\frac{\dot c}{8\pi G_D r^n},$$
which is equivalent to
$$\rho \; \eta_0\eta^1=-\frac{\dot c}{8\pi G_D r^n}.$$
Finally, we write 
$$ \rho \; g_{00}\eta^0\eta^1=-\frac{\dot c}{8\pi G_D r^n}.$$
Now eqs. (46) and (47) give
\begin{equation}
 \eta^1= \left[\frac{c'(1-\frac{c}{r^{n-1}})}{8\pi \rho G_D r^n}\right]^{1/2}. 
\end{equation}
\begin{equation}
 \eta^0= -\frac{c'}{\dot c}\left[\frac{c'(1-\frac{c}{r^{n-1}})}{8\pi \rho G_D r^n}\right]^{1/2}.
\end{equation}

\vspace{0.5cm}

\section{  Solution of A  Higher Dimensional Radiating Star}

We assume that at an instant of time $t = t_o$,  a higher dimensional relativistic spherically symmetric star starts to radiation after being born. As the time elapsed,  the
thickness of radiation zone gradually increases at the expanse of the material contents of such  star. Therefore,  the
radiation zone will be separated out from  the internal core of the star to the outside empty space. In the
internal core of the star the line-element is described by a  higher dimensional metric  given by eq.(26) and beyond the bounding
sphere of radiation zone the space-time metric is assumed to be satisfied  by the metric given by eq.(27).\\
We assume that the radius of the bounding sphere at time $t(>t_0)$ is $r=R(t)$,where the value of $c$ is $C$ and
$ M = \left[\frac{nA_nC}{16\pi G_D}\right]$.
Let us again denote 
\begin{equation} 
\Psi(c,r) = \psi(t). 
\end{equation}
Continuity of the metric component $g_{\mu\nu}$ at $r=R$ gives
$$\Psi(C,R) = \psi(t).$$
Differentiating the above relation w.r.t. time ($t$) we get,
\begin{equation}
\frac{\partial \Psi}{\partial R}\dot R = -\xi(C)\frac{\partial \Psi}{\partial C}. 
\end{equation}
Here we use the notation 
that at $r=R$, $\dot c = -\xi(C)$ and  $c'$ is almost equal to the $-\dot c$.
We now have 
$$\xi(c)= c'\left(1-\frac{c}{r^{n-1}}\right)$$
which reduces to 
$$ \xi(c)= \frac{\partial c}{\partial r}\left(1-\frac{c}{r^{n-1}}\right).$$
Finally at the position $r =  R$ we can write
\begin{equation}
 -\xi(C)\frac{\partial \Psi}{\partial C}= \frac{\partial \Psi}{\partial R}\left(1-\frac{C}{R^{n-1}}\right).
\end{equation}
In the above  $\frac{\partial \Psi}{\partial C}$, $\frac{\partial \Psi}{\partial R}$ and $\xi(C)$ denotes respectively
the values of $\frac{\partial \Psi}{\partial c}$, $\frac{\partial \Psi}{\partial r}$ and $\xi(c)$ at $r=R$.
Comparing eqs. (51) and (52) we get,
\begin{equation}
\dot R = \left(1-\frac{C}{R^{n-1}}\right)
\end{equation}
which is a first order differential equation in $R$ and can be solved. On integrating eq. (53) we get
\[
R - \frac{C^{\frac{1}{n-1}}}{n-1}\left[\ln \left(R-C^{\frac{1}{n-1}}\right)+2
\sum_{k=o}^{\frac{n-4}{2}}P_k\cos
\left(\frac{2k+1}{n-1}\right)\pi+2\sum_{k=o}^{\frac{n-4}{2}}Q_k\sin \left(\frac{2k+1}{n-1}\right)\pi\right]
\]
\begin{equation}
 =t + constant
\end{equation}
where k is positive number. We obtain the following two cases : 

(i) For  even number of dimensions ($n = D - 2$),   \\
 $$ P_k = \frac{1}{2}ln\left(x^2+2x \cos\left(\frac{2k+1}{n-1}\right)\pi + 1 \right), $$
 
$$Q_k = tan^{-1} \left(\frac{x+\cos\left(\frac{2k+1}{n-1}\right)\pi}{\sin \left(\frac{2k+1}{n-1}\right)\pi}\right).$$ 

(ii)   For odd number of dimensions  ($n = D - 2$), \\
\[
 R - \frac{C^{\frac{1}{n-1}}}{n-1}\left[ln\left[\frac{R+C^{\frac{1}{n-1}}}{R-C^{\frac{1}{n-1}}}\right]
-2\sum_{k=o}^{\frac{n-3}{2}}P_k\cos\left(\frac{2k\pi}{n-1}\right)+2\sum_{k=o}^{\frac{n-3}{2}}
Q_k\sin \left(\frac{2k\pi}{n-1}\right)\right]
\]
\begin{equation}
 =t+ constant
\end{equation}
where 
$$ P_k = \frac{1}{2}\ln \left(x^2-2x \cos\left(\frac{2k\pi}{n-1}\right) + 1 \right), $$  
$$Q_k = tan^{-1} \left(\frac{x-\cos\left(\frac{2k\pi}{n-1}\right)}{\sin \left(\frac{2k\pi}{n-1}\right)}\right). $$
Thus it is evident that if the space-time dimensions is taken more than four, one obtains a stellar solution with a radial behaviour of a radiating star  different 
from that obtained in the usual 4 dimensions.

\vspace{0.5cm}

{\bf 4. Discussion }\par

In this paper we extend the work of  Vaidya on  a radiating star  in higher dimensions.
It is found that the solution obtained in a higher dimensional framework  is different from that in  four dimensions. But we recover Vaidya solution [18]  for  $ n=2$ i.e., $D = 4$.
 Earlier Iyer and Vishveshwara [8] generalized Vaidya solution in higher dimensions  but we obtain here a  solution different from them.  For  $D > 4$, we note that 
the radius of the bounding zone of  a radiating star oscillates, which is a new result in Higher dimensions.
The period of oscillations of the radial zone is  determined
by two parameters $k$ and $n$ respectively. Thus it is evident that space-time dimension if more than four is taken then it contributes to the period of such oscillation. The other contribution on such oscillation is determined from the  mode of oscillations.
As the star radiates the radiation zone progresses but the radial size of the given star oscillates. This result reflects the  effect of spacetime dimensions on the radius of a radiating star or consequently on the size of a shining star which  may brings out information whether we are living in a universe with space-time dimensions more then four or not ?

\vspace{1.0cm}

{\it Acknowledgement} 

The authors  acknowledge {\it IUCAA Reference Centre} and {\it Physics Department} at North Bengal University for extending facilities to initiate the work.
The work is supported by UGC Research Project grant -2006. BCP  is grateful to  Profs.  N. Dadhich and A. K. Kembhavi of Inter-University Centre for Astronomy and Astrophysics ({\it IUCAA}), Pune for encouragement and fruitful discussions during a visit to IUCAA.

\pagebreak

\end{document}